\documentclass{ws-p8-50x6-00}

\usepackage{amssymb,amsmath}

\def\Journal#1#2#3#4{{#1} {\bf #2}, (#4) #3}

\def\NIMA{{\em Nucl. Instrum. Methods} A}
\def\NIMB{{\em Nucl. Instrum. Methods} B}

\def\NPA{{\em Nucl. Phys.} A}

\def\PLB{{\em Phys. Lett.}  B}
\def\PRL{\em Phys. Rev. Lett.}

\def\epp{\varepsilon^{\prime}}
\def\vep{\varepsilon}
\def\epe{\epp/\vep}

\def\pip{\pi^+}
\def\pim{\pi^-}
\def\piz{\pi^0}

\def\be{\begin{equation}}
\def\ee{\end{equation}}
\def\bea{\begin{eqnarray}}
\def\eea{\end{eqnarray}}

\newcommand{\Kl}{$K_L$}
\newcommand{\Ks}{$K_S$}
\newcommand{\Ree}{Re($\epe$)}

\begin{document}

\title{Measurement of direct CP violation by the NA48 experiment at CERN}

\author{T. Gershon
\thanks{on behalf of the NA48 Collaboration: Cagliari, Cambridge, CERN, Dubna, Edinburgh, Ferrara, Firenze, Mainz, Orsay, Perugia, Pisa, Saclay, Siegen, Torino, Warszawa, Wien}
\address{Cavendish Laboratory, Madingley Road, Cambridge, UK}}


\maketitle

\abstracts{The NA48 experiment at CERN has performed a 
measurement of direct CP violation in the 
neutral kaon system, based on data collected in 1997 and 1998. 
The preliminary result for the parameter 
$\Re(\epsilon^{\prime}/\epsilon)$ is $(14.0 \pm 4.3) \times 10^{-4}$.}

\section{Introduction}

CP violation occurs in the Standard Model through 
the imaginary phase in the CKM mixing matrix.  An alternative mechanism,
proposed shortly after the discovery of the effect\cite{disc}, 
is a superweak interaction\cite{wolf}.  The latter can be ruled out 
by experimental observation of direct CP violation, parametrized by $\epp$.
Theoretical calculations of $\epp$ within the Standard Model are hard, 
but most predictions give $\epe\sim\mathcal{O}(10^{-6})$ \cite{fab}.

The previous generation of experiments to measure direct CP violation 
(NA31\cite{na31} at CERN and E731\cite{e731} at Fermilab)
gave inconclusive results.  
New experiments were setup to clarify the situation. 
In 1999 both new experiments presented results based on the first
set of their statistics: 
Re($\epe$)=$(28.0\pm4.1)\times10^{-4}$ (KTeV, FNAL)\cite{ktev} and 
Re($\epe$)=$(18.5\pm7.3)\times10^{-4}$ (NA48, CERN)\cite{na48}.
The existence of direct CP violation is thus confirmed.
The final results of these two experiments, with substantially
smaller uncertainties, are expected to conclude on the size of
the direct CP violation effect. 

The experimental determination of \Ree\  is based on the fact that the
two CP violating neutral kaon decay amplitudes into two pions 
\begin{equation}
\begin{array}[h]{lcr}
\eta_{+-} \equiv \frac{A(K_L \rightarrow \pi^+ \pi^-)}
                      {A(K_S \rightarrow \pi^+ \pi^-)}
          \simeq \vep + \epp
& \hspace{3mm} &
\eta_{00} \equiv \frac{A(K_L \rightarrow \pi^0 \pi^0)}
                      {A(K_S \rightarrow \pi^0 \pi^0)}
          \simeq \vep - 2\epp
\end{array}
\end{equation}
contain different admixture of the two CP violating
processes in charged and in neutral mode. Therefore a double ratio
\begin{equation}
R = \frac{\Gamma(K_L \rightarrow \pi^0 \pi^0)}
         {\Gamma(K_S \rightarrow \pi^0 \pi^0)}/
    \frac{\Gamma(K_L \rightarrow \pi^+ \pi^-)}
         {\Gamma(K_S \rightarrow \pi^+ \pi^-)}
\end{equation}
is an observable sensitive to $\epe$ via $\mathrm{Re}(\epe) \simeq \frac{1}{6}(1 - R)$.

This paper describes the preliminary result from 
the analysis of data taken by NA48 in 1998. 

\section{NA48 method}
\label{sec:meth}

To measure \Ree\  to an accuracy of  $\mathcal{O}(10^{-4})$,
high statistics and good controls of systematic biases are required.
NA48 uses nearly collinear simultaneous \Ks\  and \Kl\  beams
for maximum benefit from cancellations in the double ratio.
By weighting \Kl\ decays to the \Ks\ lifetime distribution, 
performing the analysis in bins of kaon energy
and collecting all four modes simultaneously,
effects due to detection efficiencies and accidental activity are minimised. 
Low backgrounds are obtained by using high resolution detectors.  

\section{Experimental setup}

\begin{figure*}[htb]
  \begin{center}
    \includegraphics[width=9cm,height=5cm]{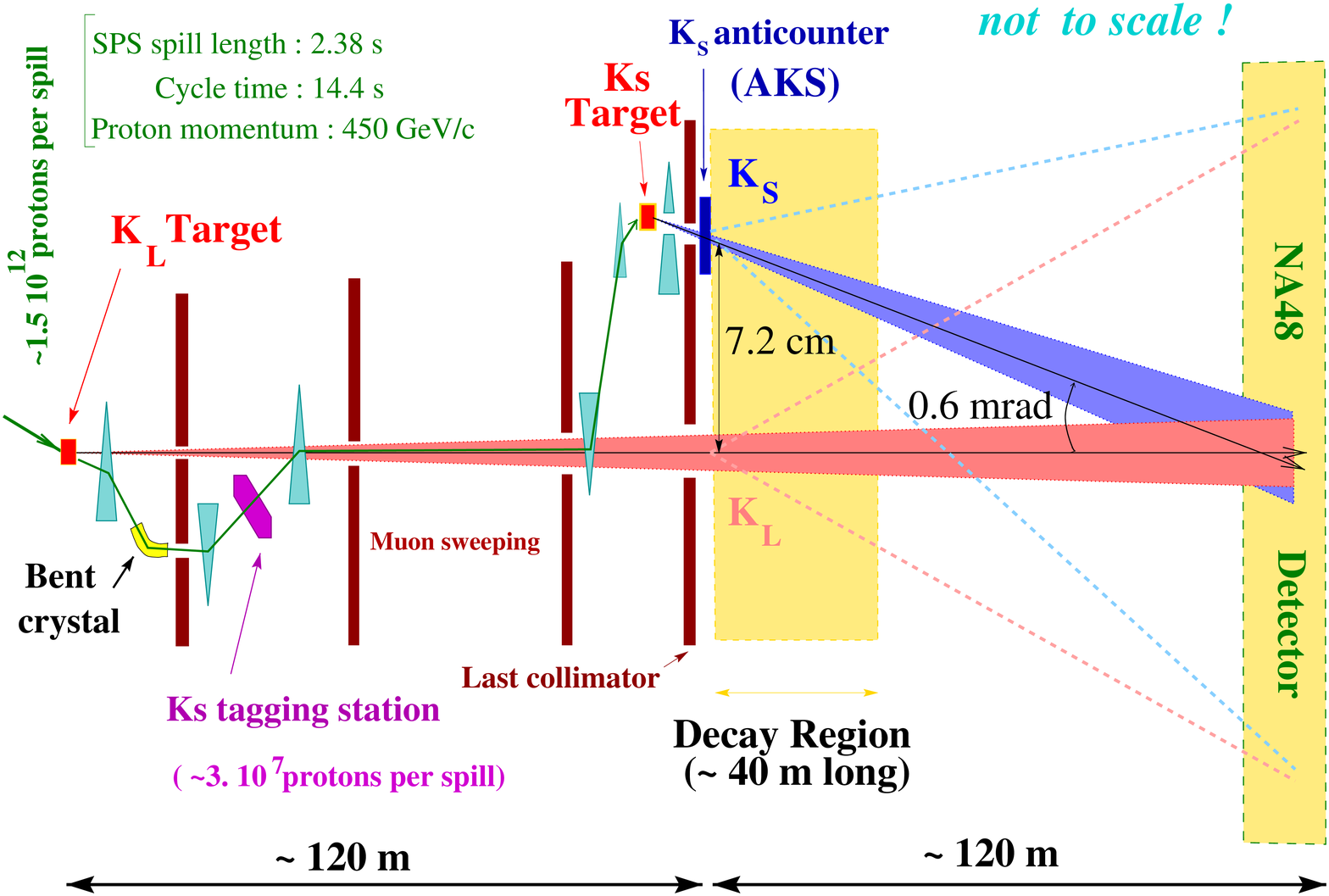}
    \caption{The layout of the NA48 experiment (side view).}
    \label{fig:experiment}
  \end{center}
\end{figure*}

The NA48 beams and detector are described in detail in \cite{nimpaper}.
A schematic is shown in figure~\ref{fig:experiment}, and the key feaures
are listed below.
\begin{itemize}
\item A primary beam of 450 GeV protons ($\sim1.4\times10^{12}$ ppp)
  is delivered by the SPS accelerator to the \Kl\ target.  A bent crystal 
  deflects the required small fraction ($\sim\mathcal{O}(10^{-5})$) 
  of the non-interacting protons through a system of tagging counters\cite{tagg}
  and onto the \Ks\ target.
\item The $\piz\piz$ decays are reconstructed using a liquid
  krypton  electro-magnetic calorimeter ($\frac{\sigma_E}{E} 
  = [0.5 \oplus \frac{3.2}{\sqrt{E/{\rm GeV}}} 
  \oplus \frac{10.}{E/{\rm GeV}}]\% $).
  The neutral trigger\cite{nut} uses calorimeter
  information and a look-up table to make a fast decision.
  The inefficiency for 2$\pi^0$ decays is $\sim$0.1 \%.
\item A magnetic spectrometer detects $\pip\pim$ decays.
  The momentum resolution is $\frac{\sigma_P}{P} = [0.5 \oplus 0.009 P({\rm GeV}/c)]\%$.
  The charged trigger consists of a fast pretrigger
  and a processor farm\cite{mbx} which computes the decay vertex
  position and invariant mass from the drift chamber signals. 
  This trigger has an inefficiency of $\sim$2.5\%,
  and dead time of $<$5 \%.
\end{itemize}

\section{Data analysis}

\subsection{Event selection and backgrounds}
\begin{figure}[htb]
\parbox{0.48\textwidth}
{
\begin{center}
    \includegraphics[width=0.45\textwidth,height=0.37\textwidth]{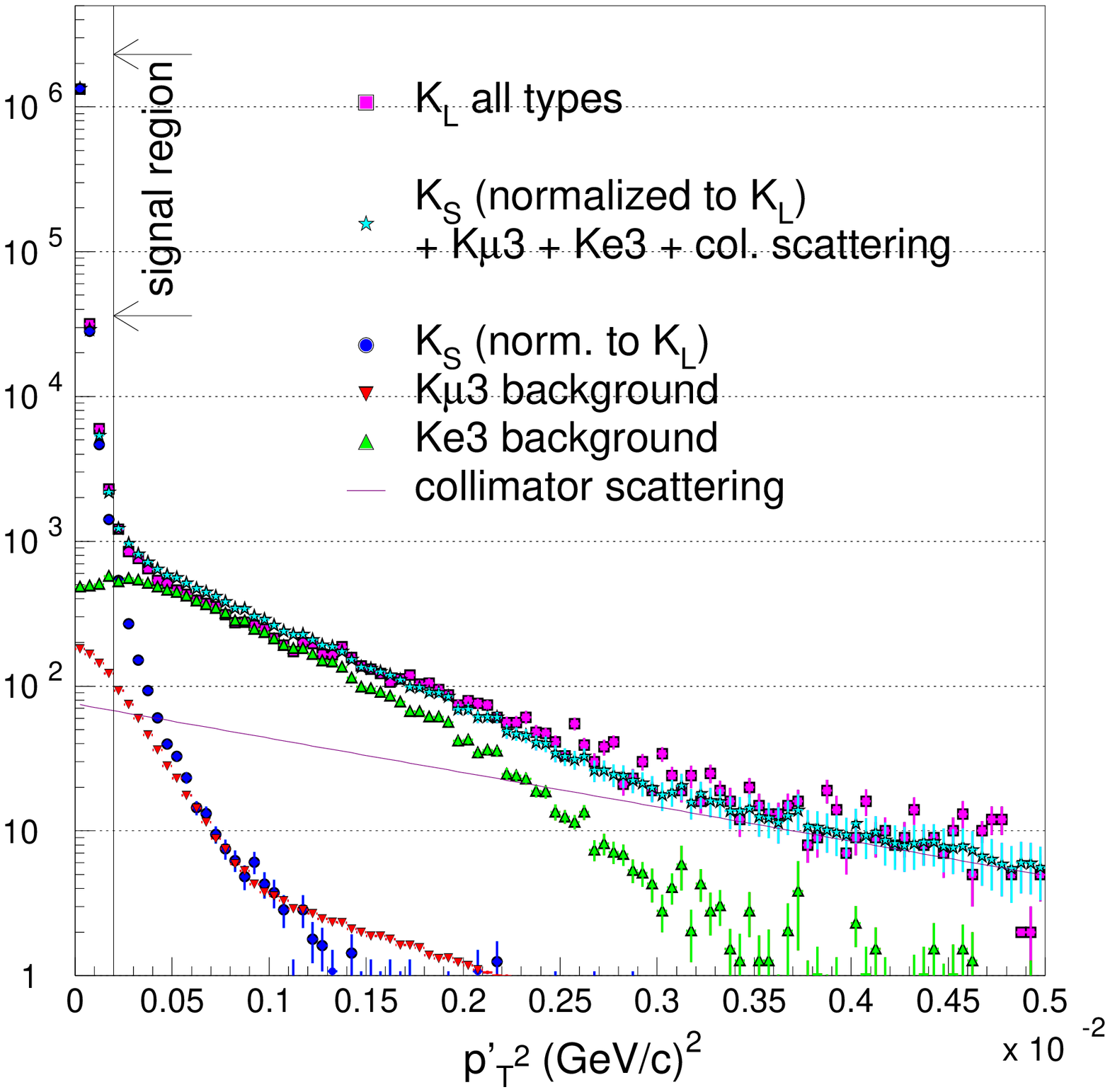}
\end{center}
}
\parbox{0.48\textwidth}
{
\begin{center}
    \includegraphics[width=0.45\textwidth,height=0.37\textwidth]{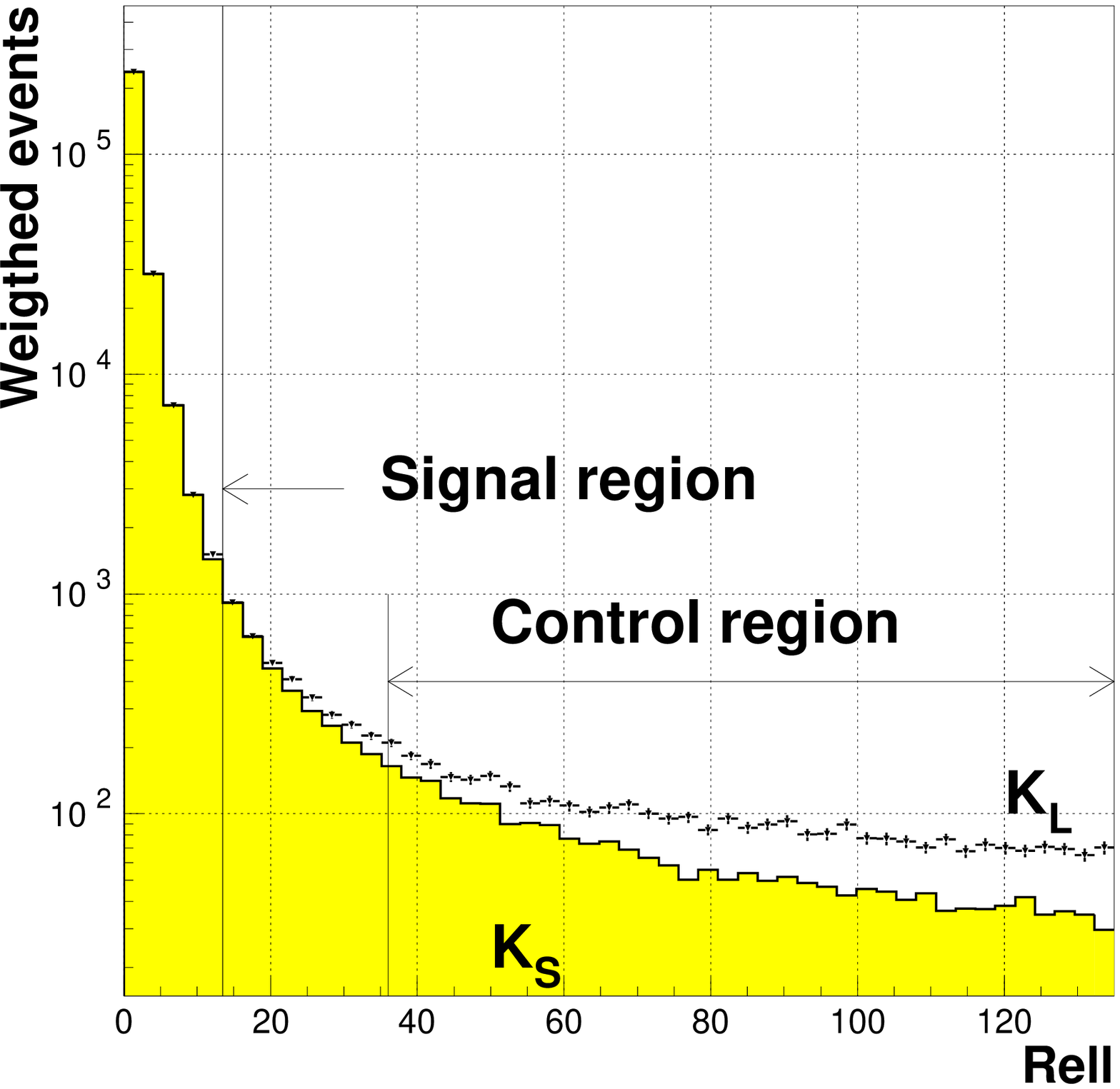}
\end{center}
}
\caption{(left) Distribution of charged signal and background components in the rescaled transverse momentum squared (${p_T'}^{2}$); (right) Distribution of neutral signal and background components in the $\chi^2$-like variable $R_{ellipse}$.}
\label{fig:bkgds}
\end{figure}
\begin{itemize}
\item All four modes are counted in the same kaon energy interval (70
  to 170 GeV), and decay volume (0 to 3.5 \Ks\ lifetimes).
  The beginning of the \Ks\  decay volume is
  determined by an anti-counter placed in the \Ks\  beam. 
\item Dead time in the trigger or read out is applied to all four modes.
\item For charged events, the momentum asymmetry rejects background from decays
  of $\Lambda$ particles.  Cuts on $E/p$ and associated muon counter hits
  reject the semi-leptonic backgrounds.  
  A signal region in invariant mass and rescaled transverse momentum squared is used.
  The background is estimated using 
  cleanly identified background decays (figure~\ref{fig:bkgds}).
\item For neutral events no additional clusters are allowed. The four clusters must
  have a low $\chi^2$ for a $K^0 \rightarrow 2\ \pi^0 \rightarrow 4\ \gamma$ hypothesis.
  The background is measured from the tail of this distribution (figure~\ref{fig:bkgds}).
\end{itemize}

\subsection{\Ks\  tagging}
\begin{figure}[htb]
\parbox{0.60\textwidth}
{
\begin{center}
    \includegraphics[width=0.45\textwidth,height=0.37\textwidth]{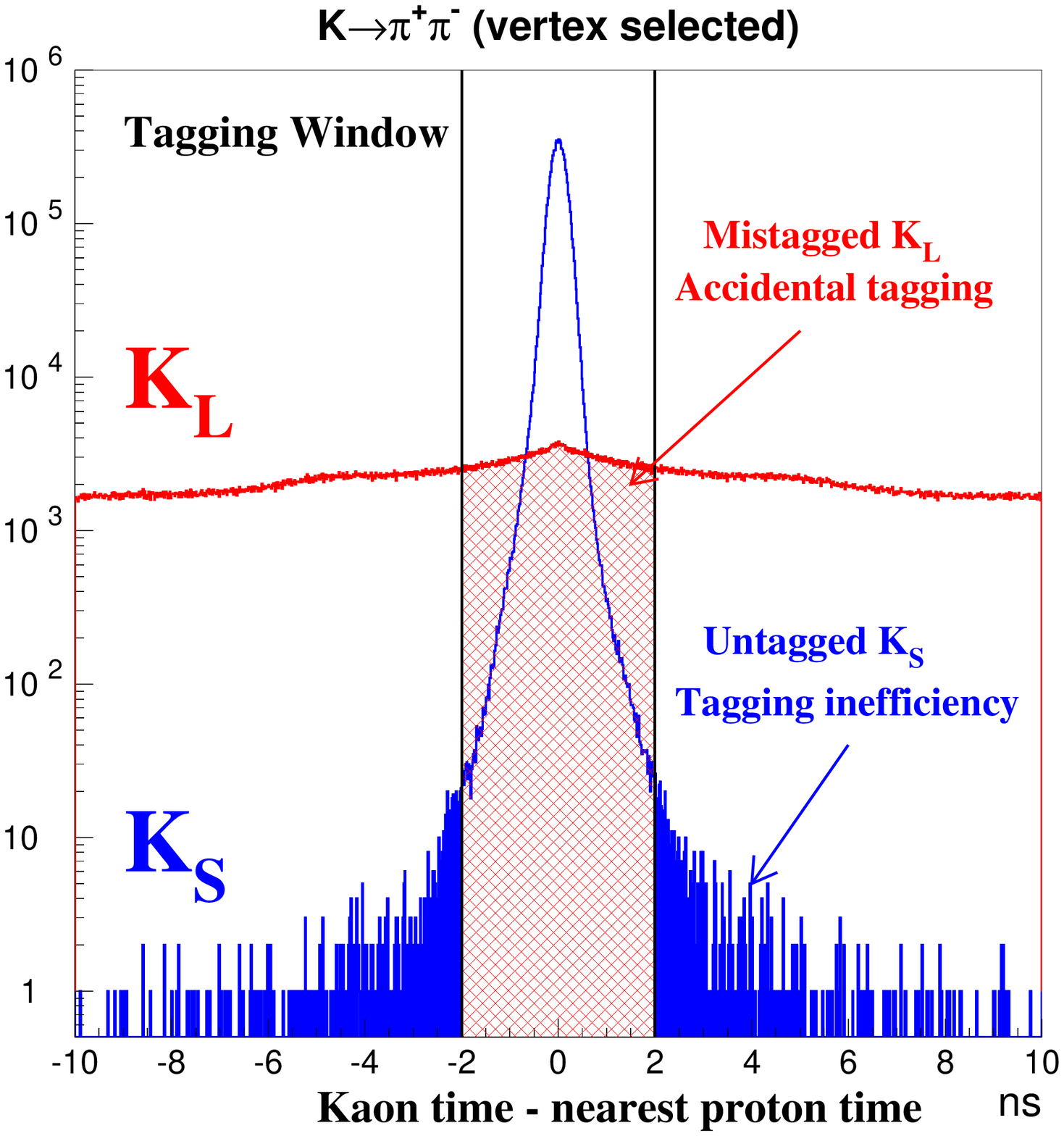}
\end{center}
}
\parbox{0.35\textwidth}
{
\begin{center}
$\alpha_{SL}^{+-}=(1.97 \pm 0.05)\times 10^{-4}$
$\alpha_{LS}^{+-}=(11.05 \pm 0.01)\%$
$\left| \Delta \alpha_{SL} \right| < 0.5\times10^{-4}$
$\Delta \alpha_{LS} = (0.3 \pm 4.2)\times10^{-4}$
\end{center}
}
  \caption{The time difference between $K \rightarrow \pi^+\pi^-$
    candidates and the nearest proton time detected by the
    tagging counter. The \Kl\  and \Ks\  contributions are
    identified by vertical vertex separation.  On the right the 
    calculated tagging quantities are given.}
  \label{fig:tag}
\end{figure}

Decays from the \Ks\ beam are identified by virtue of their having a 
coincidence between the event time and the nearest proton time 
as shown in figure~\ref{fig:tag}.
There are two forms of mistagging: 
{\it tagging inefficiency} ($\alpha_{SL}$) - at least one of the
two times was mismeasured, and 
{\it accidental tagging} ($\alpha_{LS}$) - there was an accidental
coincidences between a proton and a \Kl\ decay.  The double ratio is
sensitive only to differences between the mistagging probabilities:
\begin{equation}
\begin{array}{lcr}
\Delta R_{SL} \simeq 6 (\alpha_{SL}^{00} - \alpha_{SL}^{+-}) = 6 \Delta \alpha_{SL}
& \hspace{3mm} &
\Delta R_{LS} \simeq 2 (\alpha_{LS}^{00} - \alpha_{LS}^{+-}) = 2 \Delta \alpha_{LS}
\end{array}
\label{eq:tag}
\end{equation}
$\Delta \alpha_{SL}$ is measured using 3$\pi^0$ decays with one photon conversion, 
The value of $\Delta \alpha_{LS}$ is estimated using 
sidebands away from the coincidence peak.

\subsection{Acceptance and proper time weighting}
\label{sec:accep}

To avoid a large acceptance correction on the double ratio 
due to the difference between the two neutral kaon lifetimes,
the \Kl\  candidates are weighted with a factor to make the lifetime 
distributions the same.
After weighting the acceptance correction reduces to a value of
$\Delta {\rm R} = (+31 \pm 6_{MCstat} \pm 6_{syst})\times 10^{-4}$,
with an increase of statistical error of $\sim$35~\%.  

\subsection{Other systematics}

In order to avoid strong
sensitivity to the distance scale uncertainty the beginning of the
decay region is defined by an anti-counter, placed in the \Ks\  beam.
In the neutral decay mode the distance scale is directly related to
the energy scale. The total uncertainty on the double ratio 
from neutral reconstruction systematics is $<10 \times 10^{-4}$.

Care has to be given to losses and gains in the
event counts due to accidental activity. 
The correction on the double ratio due to different \Ks/\Kl\ 
illumination of the detector (calculated by overlaying data with
events triggered proportionally to the beam intensity)
is $(2 \pm 6) \times 10^{-4}$.
Other effects (variation of the \Ks/\Kl\ intensity ratio, noise)
are absorbed in a limit of $<10 \times 10^{-4}$.

\section{Result}

Table~\ref{tab:syst} shows the statistics collected in 1998 run,
next to a list of all corrections applied to the raw double ratio 
and of all systematic uncertainties.
\begin{table}[htb]
\parbox{0.48\textwidth}
{
\begin{center}
\begin{tabular}{c|c} \hline
Mode & Statistics/$10^6$ \\ \hline
$K_S \rightarrow \pip \pim$ & 7.5 \\
$K_L \rightarrow \pip \pim$ & 4.8 \\
$K_S \rightarrow \piz \piz$ & 1.8 \\
$K_L \rightarrow \piz \piz$ & 1.1 \\ \hline
\multicolumn{2}{c}{$\Rightarrow$}\\
\multicolumn{2}{c}{Statistical error on R : 17.3$\times$10$^{-4}$}\\
\end{tabular}
\end{center}
}
\parbox{0.48\textwidth}
{
\begin{center}
\begin{tabular}{c|r@{$\pm$}r} \hline 
Source                       & \multicolumn{2}{c}{$\Delta {\rm R}/10^{-4}$} \\ \hline
{\small Charged trigger    } & {\small  -1} & {\small 11} \\
{\small Accidental tagging } & {\small  +1} & {\small  8} \\
{\small Tagging efficiency } & {\small   0} & {\small  3} \\
{\small Neutral rec. syst. } & {\small   0} & {\small 10} \\
{\small Charged vertex     } & {\small  +2} & {\small  2} \\
{\small Acceptance         } & {\small +31} & {\small  9} \\
{\small Neutral BKG        } & {\small  -7} & {\small  2} \\
{\small Charged BKG        } & {\small +19} & {\small  3} \\
{\small Beam scattering    } & {\small -10} & {\small  3} \\
{\small Accid. activity    } & {\small  +2} & {\small 12} \\ \hline 
Total                        &         +37  &         24  \\ \hline
\end{tabular}
\end{center}
}
\caption{The statistical and systematic errors in the double ratio.
  In order to obtain the 
  effect on \Ree\  the numbers must be scaled down by factor 6.}
\label{tab:syst}
\end{table}

\begin{figure}[htb]
\parbox{0.48\textwidth}
{
    \includegraphics[width=0.45\textwidth,height=0.37\textwidth]{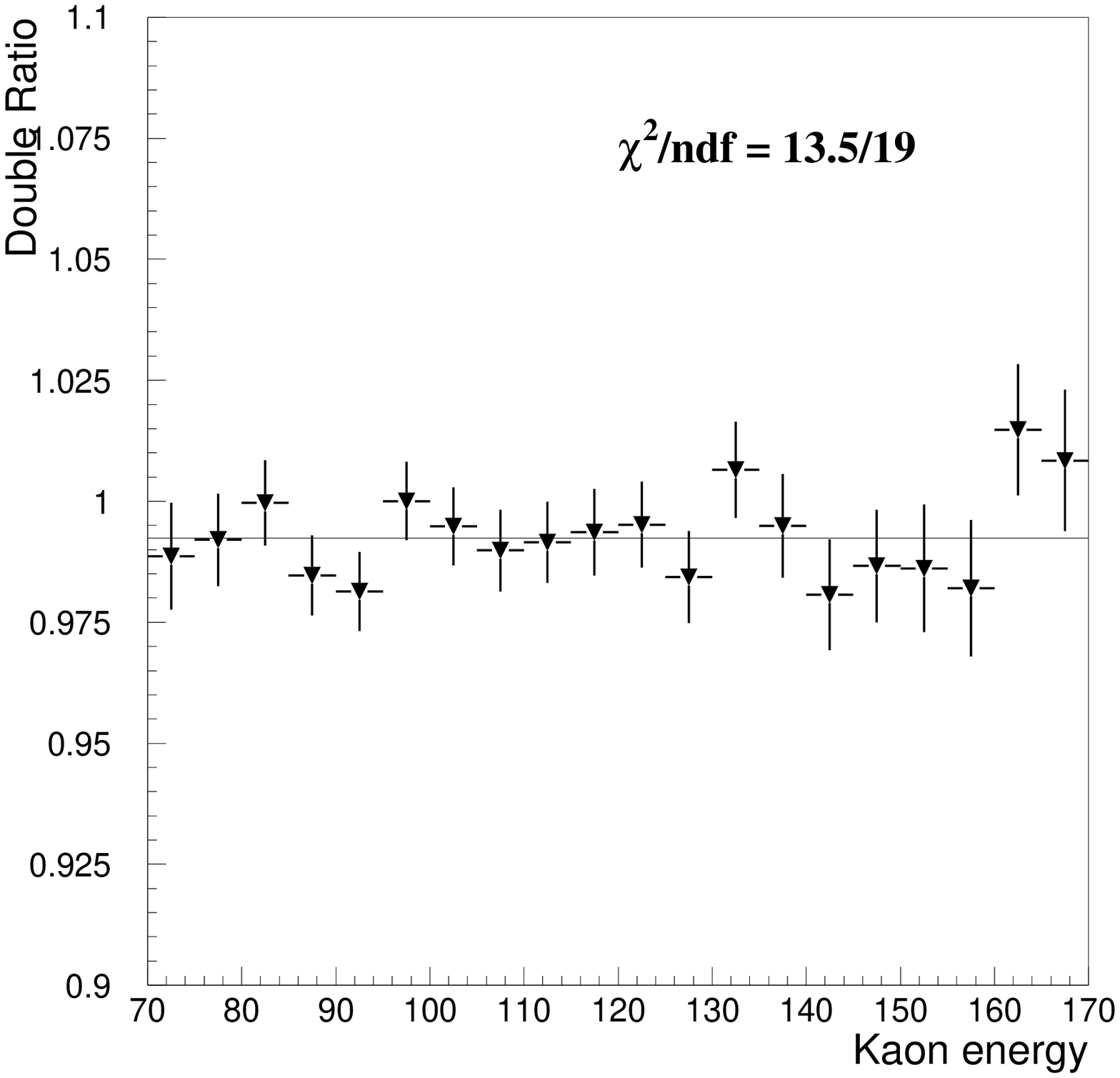}
}
\parbox{0.48\textwidth}
{
    \includegraphics[width=0.45\textwidth,height=0.37\textwidth]{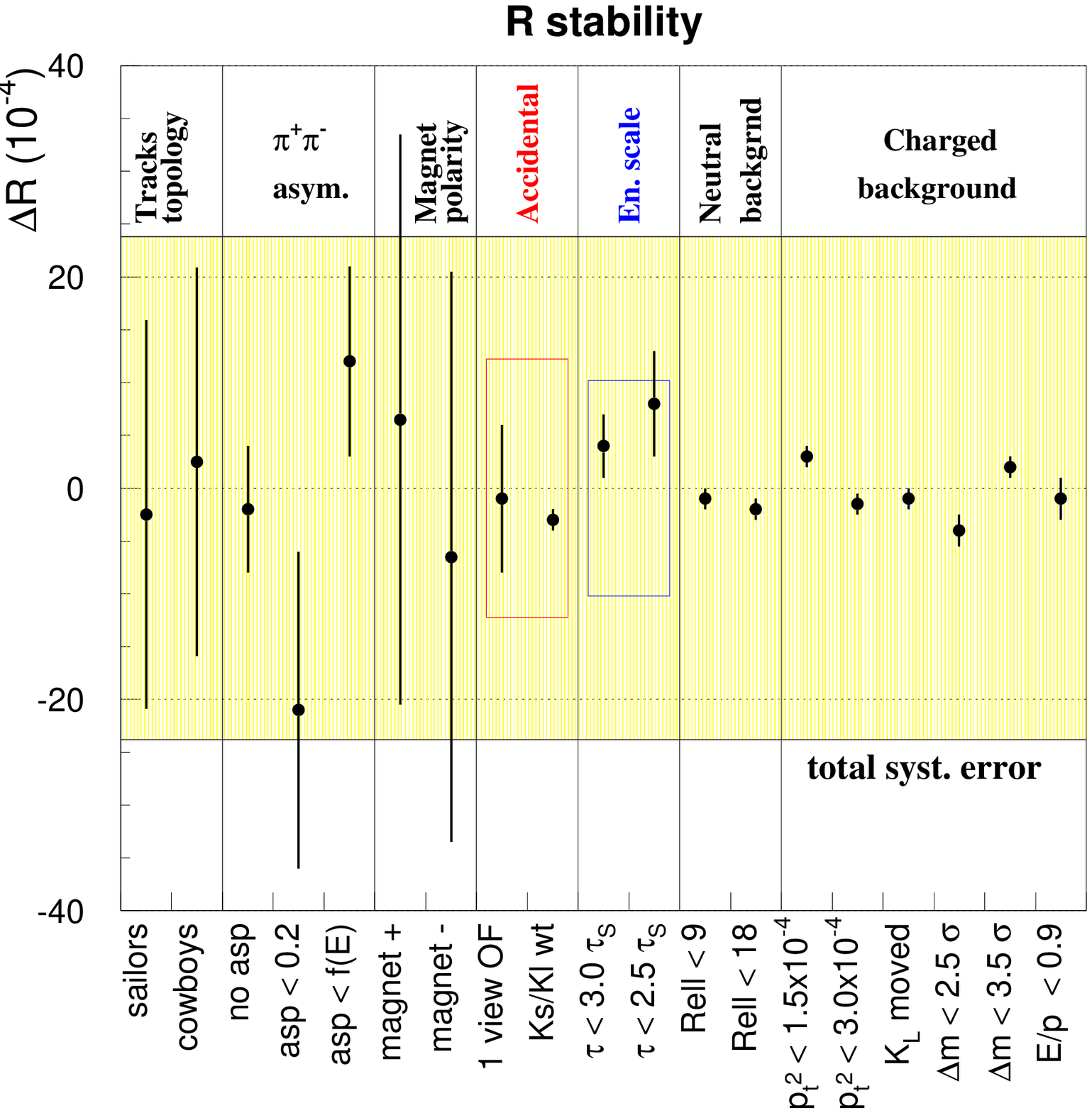}
}
\caption{The stability of the double ratio with kaon energy (left) and other checks (right).}
\label{fig:rvse}
\end{figure}

The stability of the corrected double ratio (figure~\ref{fig:rvse}) has been extensively checked.
The preliminary result obtained from the data collected in the year
1998 is
\begin{equation}
{\rm Re(}\epe{\rm)} = (12.2 \pm 2.9_{stat} \pm 4.0_{syst}) \times 10^{-4}.
\end{equation}
The combined 1997 and 1998 result is 
\begin{equation}
{\rm Re(}\epe{\rm)} = (14.0 \pm 4.3) \times 10^{-4}.
\end{equation}

\section{Conclusions and outlook}

The result presented in this paper confirms the existence of the
direct CP violation.  Further analysis on systematic studies and
the data collected in the 1999 run should lead to a total
uncertainty on \Ree\ of $\lesssim 3 \times 10^{-4}$. 

Taking into account the three most precise published 
results\cite{na31,e731,ktev} and the combined NA48 result the world average
is ${\rm Re(}\epe{\rm)} = (19.3 \pm 2.4) \times 10^{-4}$.
However the $\chi^2=10.5/3$ indicates poor consistency among the
values. New results with better accuracy 
from both KTeV and NA48 as
well as first results from KLOE\cite{kloe} at DA$\phi$NE 
should help to clarify this situation.


{\small
\begin{thebibliography}{99}
\bibitem{disc} J.H. Christenson, J.W. Cronin, V.L. Fitch and
               R. Turlay, \Journal{\PRL}{13}{138}{1964}.

\bibitem{wolf} L. Wolfenstein, \Journal{\PRL}{13}{562}{1964}.

\bibitem{fab}  S. Bertolini, J.O. Eeg, M.E. Fabbrichesi, \Journal{\em Rev. Mod. Phys.}{72}{65}{2000}.

\bibitem{na31} G.D. Barr et al., \Journal{\PLB}{317}{233}{1993}.

\bibitem{e731} L.K. Gibbons et al., \Journal{\PRL}{70}{1203}{1993}.

\bibitem{ktev} A. Alavi-Harati et al., \Journal{\PRL}{83}{22}{1999}.

\bibitem{na48} V. Fanti et al., \Journal{\PLB}{465}{335}{1999}.

\bibitem{nimpaper} The NA48 Experiment. In preparation.

\bibitem{beam} N. Doble et al., \Journal{\NIMB}{119}{181}{1996}.

\bibitem{tagg} P. Grafstr\"{o}m et al., \Journal{\NIMA}{344}{487}{1994}.

\bibitem{nut}  G. Fischer et al., \Journal{\NIMA}{419}{695}{1998}.

\bibitem{mbx} S. Anvar et al., \Journal{\NIMA}{419}{686}{1998}.

\bibitem{kloe} M. Adinolfi et al., \Journal{\NPA}{663}{1103}{2000}.


\end{thebibliography}
}
\end{document}